\documentstyle[11pt,cs11,html,epsf]{article}

\begin{document}

\title{Spectral Line Variability in the Circumstellar Environment of the Classical T Tauri Star SU Aurigae}

\author{J.M.~Oliveira\altaffilmark{1} and B.H.~Foing}
\affil{ESA Space Science Department, ESTEC, The Netherlands}
\author{ Y.C.~Unruh}
\affil{Institut f\"{u}r Astronomie, Universit\"{a}t Wien, Austria}

\altaffiltext{1}{Centro de Astrof\'{\i}sica da Universidade do Porto, Portugal}

\index{T Tauri}
\index{Cross-correlation}
\index{Balmer lines}
\index{Helium lines}
\index{Sodium lines}
\index{Magnetospheric models}

\begin{abstract}

\noindent SU Aurigae is a classical T Tauri star of spectral type G2.
This star was one of the scientific targets of the MUSICOS 96 multi-site campaign that provided a wealth of high resolution cross-dispersed spectral data, with a good continuous time coverage (Unruh et al.~1998a). We present the results of the analysis of the complex circumstellar 
environment of this star, with particular regard to magnetospheric models, in which the accretion from the disk is channelled onto the star along magnetic field lines. The signatures of modulated outflows and mass accretion events are present in the spectra,
as well as transient spectral features. We computed auto-correlation and cross-correlation functions to better investigate the source of the profiles' variability. The comparison of the profiles of different spectral lines allows us to study the footprints of events effectively observed at different distances from the stellar surface.

\end{abstract}

\keywords{T Tauri, SU Aur, spectroscopy, magnetospheric models, circumstellar environment, cross-correlation function}

\index{*SU Aur}
\index{*SU Aurigae|see {SU Aur}}
\index{*HD 282624|see {SU Aur}}

\section{Introduction}

\noindent SU Aurigae is a G2 classical T Tauri star (cTTS). It has a projected rotational velocity of $\sim$\ 66~km~s$^{-1}$ and a rotational period of about 3~days. So far the best period estimates come from the analysis of the spectral variations of the Balmer lines. Giampapa et al.~(1993) found a period of 2.98~$\pm$~0.4~days in the blue wing of the H$\alpha$\ spectral line, later confirmed also in the red wing of H$\beta$\ by Johns \& Basri~(1995b). Petrov et al.~(1996) reported a period of 3.031~$\pm$~0.003~days measured on the redshifted variations of the Balmer lines. Our data set suggests a somewhat shorter rotational period of $\sim$~2.8~days (Unruh et al.~1998a,b). Using the H$\alpha$, H$\beta$, \ion{Na}{1}~D and \ion{He}{1}~D3 profiles, we try to disentangle the two main components that are believed to be present, namely accretion and wind signatures. From equivalent width measurements of fitted components in H$\alpha$\ and H$\beta$, Johns \& Basri~(1995b) advocated that the accretion (H$\beta$\ red absorption feature) and wind (H$\alpha$\ blue absorption feature) signatures are out of phase in SU Aur, in what they call the misaligned ``egg-beater'' model or the oblique rotator, as it is also known. This model is a  generalization of the magneto-centrifugally driven flow model of Shu et al.~(1994). According to this model, a stellar dipole field truncates the accretion disk close to the co-rotation radius. At the truncation point, the ionized disk material is loaded either onto inner closed magnetic field lines, accreting thereby onto the stellar photosphere, or it is loaded onto outer open magnetic field lines that can drive a disk-wind flow. In this context, we analyse H$\beta$, \ion{Na}{1}~D and \ion{He}{1}~D3 data sets in terms of cross-correlation techniques (expanding on the work of Johns \& Basri~(1995a) for the H$\alpha$\ spectral line). Also intriguing in our data set are two drifting features appearing in H$\alpha$\ and H$\beta$ and, in one case, also in the \ion{Na}{1}~D lines.\\

\section{Description of the Data Set}

A more detailed description of our data set can be found in Unruh et al.~(1998a), but we highlight here the most important characteristics. This data set was obtained in November~1996 during the MUSICOS~96 multi-site, multi-wavelength campaign, that involved five different observatories: Isaac Newton Telescope (INT, La Palma), Observatoire de Haute Provence (OHP, France), McDonald Observatory (USA), Xinglong Observatory (China) and Canada-France-Hawaii Telescope (CFHT, Hawaii). We obtained 126 echelle spectra spanning 8.5~days, with 3.5~days covered almost continuously. H$\alpha$, \ion{Na}{1}~D and \ion{He}{1}~D3 were observed at all five observatories while H$\beta$\ could only be observed from the INT, OHP and CFHT. Our data set differs considerably from previous SU Aur data sets, because those had typically one to two spectra a night, even though over longer time spans. Our data set, by contrast, has a finer time sampling, so we are less limited by the loss of coherency of the observed phenomena. We can hence study the timescales over which the variations of the different spectral lines are related.

\section{Cross-Correlation Analysis}

\noindent The broad spectral coverage of our data set allows the comparison of lines that, due to their different ionization potentials, probe different parts of the circumstellar material. We compute the cross-correlation function of pairs of spectral lines as a function of the time lag ($\Delta$t). The data sets are interpolated to account for the unequal spacing (White \& Peterson~1994). In simple terms each velocity bin in one spectral line (at time t) is correlated with all the velocity bins of the other spectral lines (interpolated at time t+$\Delta$t) and this gives, for each pair of lines, an intensity map like the examples shown in Fig.~\ref{fig1}. These maps provide a wealth of information , even though not being necessarily the easiest to interpret. Starting with the auto-correlation function of H$\beta$ (see top row in Fig.~\ref{fig1}):

\begin{figure}
\plotfiddle{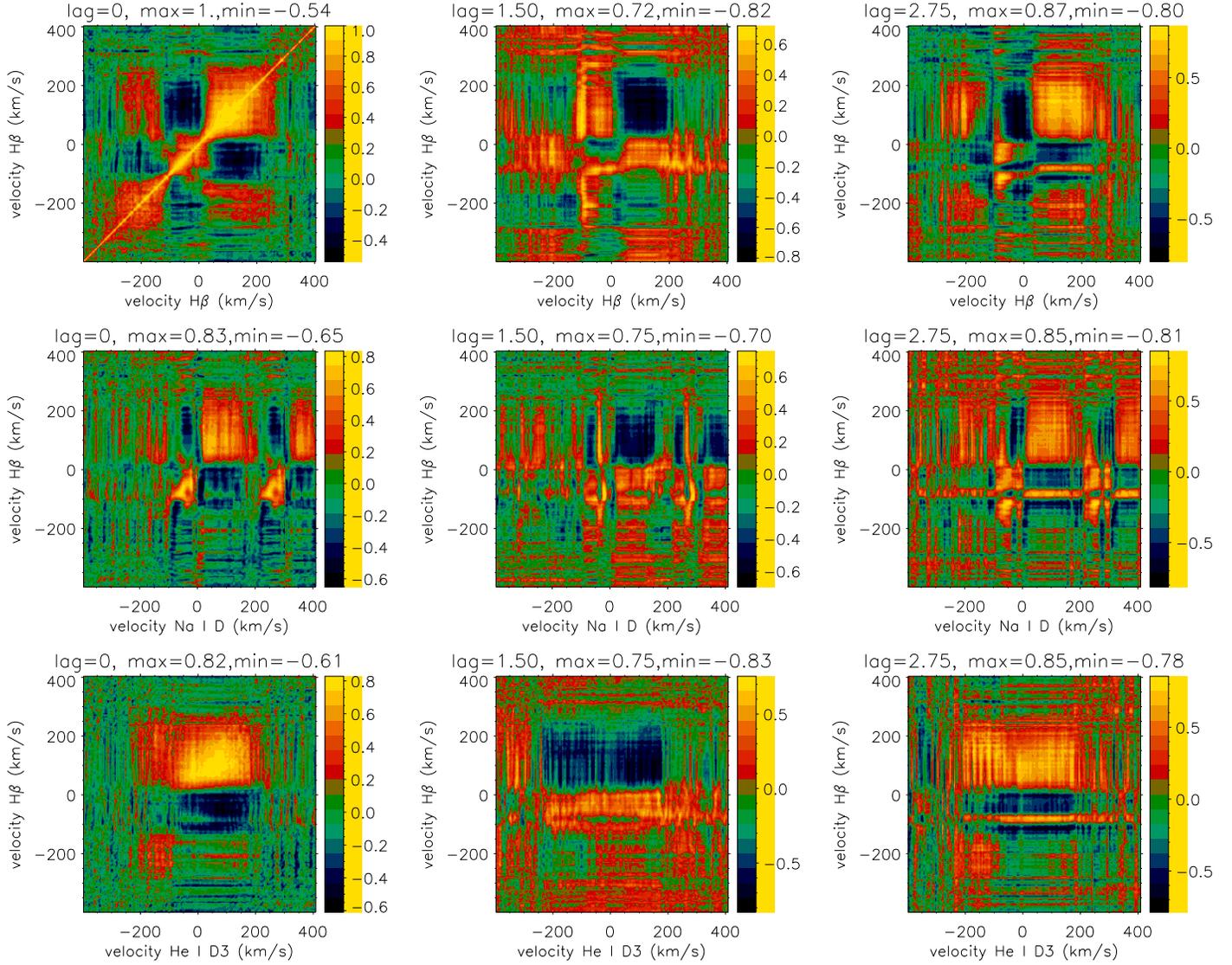}{15cm}{0}{100}{100}{-262}{0}
\caption{Cross-correlation intensity maps for H$\beta$, H$\beta$\ versus \ion{Na}{1}~D and H$\beta$\ versus \ion{He}{1}~D3. Each row represents intensity maps for one pair of lines, for 3 time lags, $\Delta$t~=~0, $\Delta$t~$\simeq$~P/2, $\Delta$t~$\simeq$~P. Yellow represents high correlation coefficients and black strong anti-correlations. The intensity (color) scale is not the same in all maps. In each map, the time lag ($\Delta$t) is indicated as well as the maximum/minimum correlation coefficients found.}
\label{fig1}
\end{figure}

\begin{itemize}
\item The red wing of H$\beta$\ auto-correlates very well over the interval [50:200]~km~s$^{-1}$.
\item The red wing of H$\beta$\ [50:200]~km~s$^{-1}$ is anti-correlated with the blue wing of H$\beta$\ [-150:0]~km~s$^{-1}$.
\item The red and blue wings are correlated albeit over a smaller velocity range for a time lag of $\sim$~1.5~days, approximately half of the detected period. For the same time lag the red wing is anti-correlated with itself.
\item With a time lag of $\sim$~2.8~days, the period we detected, we recover almost the initial cross-correlation map, due to periodic variations on both wings.
\end{itemize}

\noindent Thus the blue and red wings of H$\beta$\ seem indeed to be 180\deg\ out of phase. For the cross-correlation of H$\beta$\ and \ion{Na}{1}~D a similar analysis was performed. The resulting maps are shown in the middle row in Fig.~\ref{fig1}. The map at $\Delta$t=0 shows the same features as the H$\beta$\ cross-correlation map, although over a narrower velocity range. This suggests that the two \ion{Na}{1} resonance lines behave very similar to the H$\beta$\ line. In particular we find:
\begin{itemize}
\item the red wings of the two \ion{Na}{1}~D lines [25:150]~km~s$^{-1}$ correlate well with the red wing of H$\beta$ [50:200]~km~s$^{-1}$.
\item With a time lag of about half of the detected period, these parts of the profiles become anti-correlated.
\item There is a weak correlation between the blue wings of \ion{Na}{1}~D [-100:0]~km~s$^{-1}$ and H$\beta$ [-150:0]~km~s$^{-1}$.
\item The red wings of \ion{Na}{1}~D [25:150]~km~s$^{-1}$ are anti-correlated with the blue wing of H$\beta$ [-150:0]~km~s$^{-1}$.
\end{itemize}

\noindent Therefore the red wings of \ion{Na}{1}~D and H$\beta$\ vary approximately in phase. The case of the cross-correlation between H$\beta$\ and \ion{He}{1}~D3 is slightly different, as the whole \ion{He}{1}~D3 seems to vary in a concerted way (see bottom row in Fig.~\ref{fig1}). We find that:
\begin{itemize}
\item The whole \ion{He}{1}~D3 profile [-100:150]~km~s$^{-1}$ correlates well with the red wing of H$\beta$ [25:220]~km~s$^{-1}$; in these regions the 2.8~day periodicity was detected.
\item With a time lag of about half of the detected period the correlation diagram shows these lines as anti-correlated.
\end{itemize}

\noindent In conclusion, we find that the blue and red wings of H$\beta$\ are approximately 180\deg\ out of phase, while the red wings of \ion{Na}{1}~D and H$\beta$ vary in phase. Finally, the whole profile of \ion{He}{1}~D3 varies in phase with the red wing of H$\beta$. The time lags here indicated have to be taken with caution, as no error estimates were performed.

\section{Oblique ``Egg-beater'' Model Predictions}

\noindent Johns \& Basri (1995b) found that the blue wing of H$\alpha$\ and the red wing of H$\beta$ were approximately 180\deg\ out of phase. They proposed that this behaviour could be explained by an oblique ``egg-beater'' model (see Fig.~15 of Johns \& Basri~1995b) that predicts that the signatures of mass accretion and disk winds should be rotationally modulated and 180\deg\ out of phase, since at different phases the visible wind flow or funnel flow are favoured. From the H$\beta$ auto-correlation analysis we indeed have confirmation that the red wing and the blue wing absorptions vary in this way, meaning that the accretion and disk-wind signatures are in phase opposition. Also favouring this model is the fact that the variance profile of H$\beta$ shows in the red wing a preferred projected velocity (as expected from a funnel geometry); this is not seen in the blue wing, as this part of the profile originates in a less ``collimated'' disk-wind. The red wings of the \ion{Na}{1}~D lines vary in phase with the red wing of H$\beta$, a clear sign that they are also formed in accretion flows. Futhermore, the red wing of H$\beta$ varies in phase with the whole profile of \ion{He}{1}~D3. This can be taken as an indication that the helium line excitation is also related to the accretion but not to the wind in SU Aur. \ion{He}{1}~D3 excitation needs very high temperatures (or densities), therefore it can not be simply formed in the accreting flows. If the source of the excitation is the energy released at the ``foot-points'' of the accretion streams, then one could expect that the H$\beta$ red wing (accretion) and the \ion{He}{1}~D3 line vary approximately in phase, as we have found.\\

\section{Transient Features}

\begin{figure}[t]
\plotfiddle{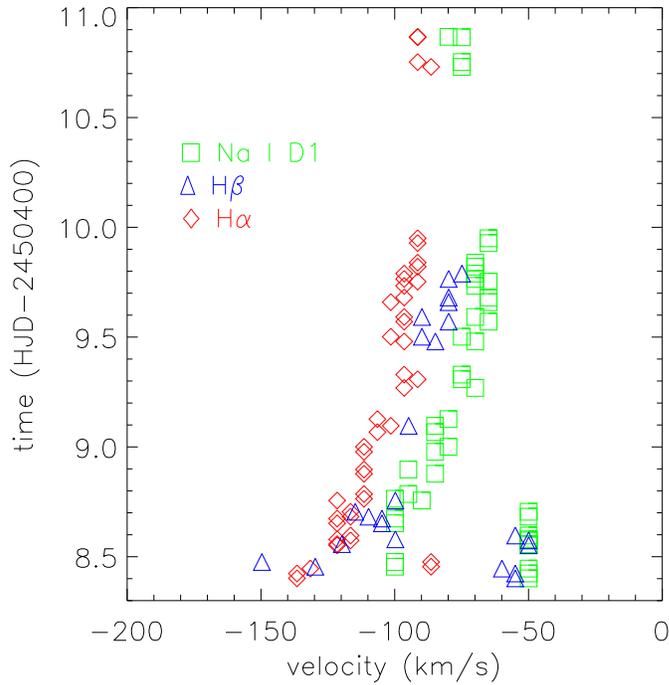}{7cm}{0}{100}{100}{-150}{0}
\caption{Velocity evolution of the drifting absorption feature seen in the H$\alpha$, H$\beta$\ and \ion{Na}{1}~D1 line profiles. For \ion{Na}{1}~D2, the velocity position of the feature is similar to \ion{Na}{1}~D1 and we do not represent it here for clarity's sake.}
\label{fig2}
\end{figure}

\noindent A drifting feature appears between the eighth and eleventh days of our observations in the profiles of H$\alpha$, H$\beta$ and \ion{Na}{1}~D (see Fig.~1 of Unruh et al.~1998b). By analysing the \ion{Na}{1}~D profiles, it is clear that this is an absorption component. As this part of the profile correlates well with the corresponding parts in the Balmer profiles, we treat it as the same drifting absorption feature, present in all three spectral lines. It is important to point out that no clear periodicity was detected in this part of the profiles for any of the lines.
We have measured the velocity position of the minima of the line profiles in the relevant velocity interval and in Fig.~\ref{fig2} we show their velocity evolution. It seems to indicate that v(H$\alpha$)~$<$~v(H$\beta$)~$<$~v(\ion{Na}{1}~D). However, the velocity measurements for H$\beta$\ have to be treated with caution as the feature can not be identified unambiguously in this line. Instead, we observe that the blueshifted absorption gets broader and deeper. The velocity difference (between the lines) probably traces material accelerating outwards. It should be stressed that approximately 1~day after the decay of the feature described above, another transient feature was detected in H$\alpha$ and H$\beta$ but no counterpart was seen in the \ion{Na}{1}~D lines.

\section{Conclusions}

\noindent The present analysis supports the so-called oblique or misaligned ``egg-beater'' model (Johns \& Basri~1995b). Our data set covers 8.5~days with several spectra each night, and at least 3.5~days of almost continuous coverage, allowing us to compare different (parts of the) line profiles and study over which time scales are they related.
The temporal relations we have found between the spectral lines help to clarify which parts of the line profiles variations originate from which component of the circumstellar environment of SU Aur: the red wing absorption components of the H$\beta$ and \ion{Na}{1}~D lines form in the accretion funnels, \ion{He}{1}~D3 forms in the highly energetic region at the ``foot-points'' of these columns, and the slightly blueshifted absorption feature in the blue wing of H$\beta$ forms in a disk-wind flow. Also present in the data set are transient features not fully understood, possibly the signature of an outwardly accelerating (stellar) wind component.

\acknowledgements{JMO acknowledges the support of the {\it Funda\c{c}\~{a}o para a Ci\^{e}ncia e Tecnologia} (Portugal) under the grant BD9577/96. YCU a\-cknow\-led\-ges the support through grant S7302 of the Austrian {\it Fond zur Wissenschaftlichen F\"{o}rderung}. The authors wish to thank the MUSICOS 96 collaboration and the staff in all the observatories involved.}

\end{document}